\documentstyle[aps,prl,epsf]{revtex}
\def\bei{\begin{itemize}}
\def\eei{\end{itemize}} 
\def\ben{\begin{enumerate}}
\def\een{\end{enumerate}} 
\def\beq{\begin{equation}} 
\def\eeq{\end{equation}}
\def\ber{\begin{eqnarray}} 
\def\eer{\end{eqnarray}} 
\def\l{\label}

\def\v#1{\vec{#1}} 
\def\b#1{\bar{#1}}

\def\Ms{M^*}
\def\Msd{$M^*$ }

\def\e{&=&}

\def\e{&=&}

\def\ls{\langle \vec{p}'\mid}
\def\rs{\mid\vec{p}\rangle}
\def\ll{\langle \vec{\ell}'\mid}
\def\rl{\mid\vec{\ell}\rangle}

\def\ubp{\bar{u}^{*}(\frac12\vec{P}+\vec{p}')} 
\def\up{u^{*}(\frac12\vec{P}+\vec{p})}
\def\ubl{\bar{u}^{*}(\frac12\vec{L}+\vec{\ell}')} 
\def\ul{u^{*}(\frac12\vec{L}+\vec{\ell})}
\def\ubmp{\bar{u}^{*}(\frac12\vec{P}-\vec{p}')} 
\def\ump{u^{*}(\frac12\vec{P}+-\vec{p})}
\def\ubml{\bar{u}^{*}(\frac12\vec{L}+-\vec{\ell}')} 
\def\uml{u^{*}(\frac12\vec{L}+-\vec{\ell})}
\def\lNM{\langle {\rm NM}\mid}
\def\rNM{\mid  {\rm NM}\rangle}

\def\ubp{\bar{u}(\vec{p},s)}

\def\up{u(\vec{p},s)}

\def\e{&=&}

\def\ls{\langle \vec{p}'\mid}
\def\rs{\mid\vec{p}\rangle}
\def\ll{\langle \vec{\ell}'\mid}
\def\rl{\mid\vec{\ell}\rangle}

\def\MEla{\frac{1}{\bar{E}(\vec{\ell}_1)}}
\def\MElb{\frac{1}{\bar{E}(\vec{\ell}_2)}}

\def\ubp{\bar{u}(\vec{p}')}
\def\up{u(\vec{p})}
\def\ubl{\bar{u}(\vec{\ell}')}
\def\ul{u(\vec{\ell})}
\def\ubmp{\bar{u}(-\vec{p}')}
\def\ump{u(-\vec{p})}
\def\ubml{\bar{u}(-\vec{\ell}')}
\def\uml{u(-\vec{\ell})}

\draft
\begin{document}
\title{\hfill DOE/ER/40762-137 \\
\hfill U.Md. PP\# 98-056 \\
The role of tensor force in nuclear matter saturation} \author{Manoj
K. Banerjee$^{1,2}$ and John A. Tjon$^{1,3}$}

\address{$^1$ Department of Physics, University of Maryland, \\
College Park, MD, USA, 20742-4111\\
$^2$ IKP, Forschungszentrum-J\"{u}lich, 52425 J\"{u}lich, Germany\\
$^3$ Institute for Theoretical Physics,
University of Utrecht \\
3508 TA Utrecht, The Netherlands}
\date{\today}

\maketitle

\begin{abstract}
Using a relativistic Dirac-Brueckner analysis the pion
contribution to the ground state energy of nuclear matter is
studied. Evidence is presented that the role of  the tensor force 
in the saturation mechanism is
substantially reduced compared to its dominant role in a usual 
non-relativistic treatment. 
The reduction of the pion contribution in nuclear matter is due 
to many-body effects present in a  relativistic treatment.  In
particular,  we   show that the damping of OPEP is actually due to the 
decrease of $M^*/M$ with increasing density.  

\end{abstract}


Within a non-relativistic framework the tensor force plays a dominant role both
in the formation of the deuteron and in nuclear matter saturation
mechanism. For the deuteron $$\langle ^3D_1\mid V_{\rm total}\mid ^3S_1\rangle
\simeq
2\langle ^3D_1\mid V_{\rm tensor}\mid ^3S_1\rangle \sim -24\,{\rm MeV},$$ while
the kinetic energy is $\sim 22\,{\rm MeV},$ adding up to a binding energy of
$\sim -2$MeV.
In a non-relativistic Bethe-Brueckner  calculation of nuclear matter one finds
typically~\cite{FG}
\beq
\lNM V_\pi\rNM_{\rm {non-relativistic}} \sim -34\,
(\rho/\rho_0)^{0.45}\,{\rm MeV}, \l{vpinr}\eeq  where $\rho_0=0.17$ fm$^{-3}$ is
the saturation density of normal nuclear matter. The
exponent of $\rho$ is  markedly less than the nominally expected value of $1$ because of
Pauli blocking. It is more than four decades ago that it was pointed out that in
the absence of the tensor force the saturation properties of normal 
nuclear matter cannot be understood~\cite{Bethe1}, assuming the 
validity of a non-relativistic quantum description. In particular,
saturation occurs   at a density one order of magnitude or more
higher  than $\rho_0$. The
effect of the pion, being given essentially by the second order in the
one-pion-exchange potential (OPEP) contribution, is attractive and
large at low density. Pauli blocking reduces the attraction  as density
increases. This mechanism leads to
saturation of nuclear matter  at a  density well
in correspondence with the empirical data. The above conclusions are
obviously  based on a non-relativistic description of the nuclear
many body system.

The relativistic results for the contribution  of $V_\pi$ to the
deuteron~\cite{HumTj} is
$\langle {\rm D}\mid V_\pi\mid {\rm  D}\rangle =-31{\rm MeV},$
suggesting an equally important role of the pion. In sharp
contrast, this seems not to be the case in a relativistic treatment of nuclear matter.
Strong scalar ($S$) and vector ($V$) fields\footnote{To be precise, here we
refer to the scalar and vector parts of the self-energy as fields.}
of the order of a few hundred MeV are typical for  relativistic
theories~\cite{mth,AmTj,Mach} based on a meson theoretical
description of the nuclear force. These values are
consistent with expectations based on the studies of scattering of $\sim
1$ GeV protons by nuclei. The large scalar fields have far reaching
consequences in nuclear matter through the strongly medium
modified nucleon mass $M^*=M+S$. The saturation mechanism is believed to
rest upon the decrease of $M^*$ with increasing
density. It is the only possible mechanism for saturation in a
mean field theory (MFT) like the QHD~\cite{SW}.
A Dirac-Brueckner (D-B) analysis~\cite{mth,AmTj} is at present
the best tool we have for a relativistic study of nuclear matter.  In Ref.~\cite{AmTj} one finds $S_{D-B}=-306\,(\rho/\rho_0)^{0.81}\,$MeV, while $V_{D-B}=233\,(\rho/\rho_0)^{0.97}\,$MeV. This reduction in the rate of increase of the strength of $S$  compared to the increase of $V$ with increasing density is the  saturation mechanism in D-B.
In a MFT the vector field  and the scalar field arise from the exchange of the $\omega$ and 
the $\sigma$ meson. A MFT treatment using the
 interaction  of Ref.~\cite{AmTj}, with only the $\sigma-$ and $\omega$-exchange contributing,  yields $S_{MFT}\sim - 358 \,(\rho/\rho_0)^{0.92}$
MeV and $V_{MFT}\sim 295$
MeV.  The reduced rate of increase of $S_{MFT}$ with increasing $\rho$  is entirely due to the decrease of scalar charge of the nucleon.  In
a D-B study, where ladders of meson  exchanges are summed, there are other
 factors contributing to both  $S$ and $V$.  As a result there are  significant differences in both magnitude and
density dependence of $S$ and $V$ in the two treatments. Thus when ladders are summed  the saturation mechanism need not be
exclusively due to $\sigma$-exchange. 

Because of the dominant role of OPEP  in
the saturation mechanism in non-relativistic studies of nuclear matter, one
should expect a similar role in relativistic studies. The important role of the 
density dependence of  $S$  does not exclude it.
In this letter we examine the role of OPEP in D-B and
show that it is substantially reduced due to relativity.  Since the contribution of OPEP to the deuteron binding energy remains large in a relativistic treatment the  damping in nuclear matter must be due to many-body effects.  We find that it can  be attributed to the decrease of $M^*/M$ with increasing density.
Assuming that  the  nuclear matter is uniform in space and constant in time
with a given density $\rho $, the
baryon current is given by $B^\mu= \rho  u^\mu$ with
$u^\mu=(0, \v 1)$ being the unit vector in the nuclear
matter frame.
Relativistic covariance implies that the
self-energy contribution of the nucleon with momentum $p$ can be characterized
by
\beq
\Sigma=\Sigma^s-\Sigma^u \gamma\cdot u -\Sigma^v \gamma \cdot  p.
\l{selfE}
\eeq
The medium modified mass of the nucleon is then given by
$$\Ms=(M+\Sigma^s)/(1+\Sigma^v).$$
Scaling all relevant momenta in terms of $\Ms$ units we may
express the ground state binding energy
$E/A$ in terms of a dimensionless NN $G$-matrix $\b G$ as
\ber
E/A\e\Ms(\frac{3\pi^2}{2\b k_F^3}) [8\int_0^{k_F/\Ms}\frac{d^3\ell}{(2\pi)^3}
(\b E(\v \ell)-1)
\nonumber\\
&&
+\frac12\sum_{(\lambda,i)}\int_0^{k_F/\Ms}\frac{d^3\ell_1}{(2\pi)^3}
\MEla
\int_0^{k_F/\Ms}\frac{d^3\ell_2} {(2\pi)^3}\MElb\langle \v
\ell_1,\lambda_1,i_1;\v \ell_2,\lambda_2,i_2\mid \b G
\mid \v \ell_1,\lambda_1,i_1; \v \ell_2,\lambda_2,i_2 \rangle],
\l{EbyA}
\eer
where $\ell_n=p_n/\Ms$ are the scaled momenta of the two
interacting nucleons and $\b E(\v l)=\sqrt{\v l^2+1}$.
The ${G}$-matrix satisfies the
Dirac-Bethe-Brueckner-Goldstone equation involving dimensionless
quantities. Within a relativistic quasipotential approach it
has in the NN c.m. system the form
\ber
\langle \v \ell'\mid && \b G \mid \v \ell\rangle = \langle \v
\ell'\mid\b V\mid\v \ell\rangle
\nonumber
\\
&& + \sum_{\rho,\lambda,i}\int\frac{d^3l''}{(2\pi)^3}
\ \ \langle \v \ell'\mid \b V\mid\v l''\rangle \ \
{\b S}_2(l'') \ \  \langle \v l''\mid\b G \mid\v \ell\rangle,
\l{BGE2}
\eer
where we sum over the $\rho$-spin, helicities and isospin
and integrate over the scaled relative momenta $\v l''$
of allowed intermediate states. Furthermore,
$S_2$ is the 2-nucleon Green function, including the
Pauli-blocking operator $\b Q_{Pauli}$.
Using a Blankenbecler-Sugar-Logunov-Thavkhelidze prescription
${\b S}_2$ has the form~\cite{BSu}
\ber
{\b S}_2(l'')&& = \frac{\pi \Ms}{(1+{\Sigma^v})}
\nonumber \\
&&\times\frac{[ E_f \gamma_0 - \v k \v \gamma+\Ms] [ E_f \gamma_0 +
\v k \v \gamma+\Ms]}{( E^* + E_f)^2 ( E^*_{k}- E_f - i \epsilon)}
{\b Q_{Pauli}}
\eer
with $k= l'' \Ms$, $E_f=W_0/(1+{\Sigma^v})$ and
$W_0^2=(p_1+p_2)^2$, $W_0$
being the total invariant energy of the final state.
Furthermore, in Eq.~(\ref{BGE2}) $\b V$ are the dimensionless
quasi-potential matrix elements in Dirac space
\ber
\langle \v \ell' \mid \b V && \mid \v \ell\rangle=
{\Ms}^2
\nonumber\\
&&\times
[ \ubp^{(1)}_{\lambda'_1,\rho'_1} \ubmp^{(2)}_{\lambda'_2,\rho'_2}
 V \up^{(1)}_{\lambda_1,\rho_1} \ump^{(2)}_{\lambda_2,\rho_2}],
\eer
where $u$'s are the positive and negative energy ($\rho=\pm 1/2$)
spinors for mass $\Ms$ fermions,  satisfying
\beq
(\rho E^* \gamma_0-\v \gamma\cdot{\v p}-\Ms)
u_{\lambda,\rho} (\v p)=0.
\l {Direq}
\eeq
with $E^*=({\v p}^2+M^{*2})^{1/2}$.
Interdependence of the nucleon self-energy  and the
$G$-matrix, for a fixed given nuclear matter density,
requires selfconsistency in solving the D-B  equations.
The NN quasipotential is taken to be given by the
relativistic one-boson-exchange (OBE) model with
$\pi, \rho, \omega, \sigma, \delta$ and $\eta$-mesons.
We have, ignoring the isospin factors, for the scalar, vector and pseudo-scalar
meson exchanges in nuclear matter
\ber
\ls V_{\sigma}\rs &=&
-g_\sigma^2 \frac{[\ubp\up]^{(1)}[\ubmp\ump]^{(2)}} {(\v p'-\v
p)^2+m_\sigma^2},
\l {scalar} \\
\ls V_{\omega,\rho}\rs &=&
g_V^2 [\ubp\{\gamma_\mu+i\frac{f_V}{2M}
\sigma_{\mu\nu}(p'-p)^\nu\}\up]^{(1)}
\frac{1}{(\v
p'-\v p)^2+m_V^2}[\ubmp\{\gamma^\mu-i\frac{f_V}{2M}
\sigma^{\mu\nu}(p'-p)_\nu\}\ump]^{(2)},
\l {vector} \\
\ls V_{P}\rs &=&
 -(\frac{g_{\pi NN}}{2M})^2  [\ubp\gamma_5
(p\!\!\!/'-p\!\!\!/)\up]^{(1)}
\frac{1}{(\v p'-\v
p)^2+m_P^2}[\ubmp\gamma_5 (p\!\!\!/'-p\!\!\!/)\ump]^{(2)} .
\l {pseudo}
\eer
We will assume, that possible density dependence
of meson-nucleon couplings and meson masses~\cite{BT} can
be neglected.

Using an angular-averaged Pauli-blocking operator we may
write for the 2-nucleon Green function
\beq
{\b S}_2 =
\frac{M^{*2}}{E^{*\,2}(\v P,\v p)-E^{*\,2}(\v P,\v k)}
\approx \frac{1}{(\v p/\Ms)^2-(\v k/\Ms)^2},
\l{gbdef}
\eeq
The last line of Eq.~(\ref{gbdef}) indicates that the range of the
intermediate momentum $\v k$ is controlled by \Msd.
The smaller the \Msd, the smaller is the contributing range
of $\v k$. This fact has an
important effect on the $\pi$ and $\eta$ and the Pauli
coupling of the vector mesons. The corresponding matrix elements have
two extra powers of $\v k$ in the numerator compared to other matrix
elements. So with increasing density and, therefore, with decreasing
\Msd these matrix elements have  suppression factors not present in all
other matrix elements.

Given the solutions of the $G$-matrix the binding energy is
calculated using Eq.~(\ref{EbyA}). The potential contributions
of the pion can be calculated using the
Hellmann-Feynman theorem
\beq
\lNM V_{\pi}\rNM=g_{\pi NN}^2\frac{\partial}{\partial \,g_{\pi NN}^2}(E/A),
\l{Vpi} \\
\eeq
and  we find, that the contribution  of $V_\pi$ to $E/A$ is
\beq
\lNM
V_\pi\rNM_{\rm Relativistic} \sim -20\,(\rho/\rho_0)^{0.16}\,{\rm MeV}.
\l {ROPEP}
\eeq
It is considerably suppressed  compared to the  value given by Eq.~(\ref{vpinr}) for the non-relativistic case.
\begin{figure}[htb]
\epsfxsize=2.5in
\epsfysize=2.5in
\vspace{0.2in}
\epsffile{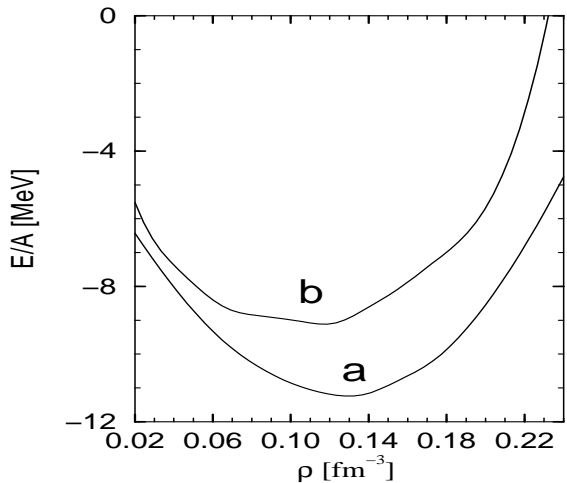}
\vspace{0.2in}
\caption[F4]{Plots of the quantities $E/A$ (curve a) and
$E/A-\langle NM\mid v_\pi\mid NM\rangle-17\,$MeV (curve b).
The numbers are based on the work   of
Ref.~\protect{\cite{AmTj}}.}
\l {novpi}
\end{figure}
Furthermore, it has only a minor role in the saturation
mechanism.  This is exhibited in Fig.~\ref{novpi} where we plot $E/A$ (curve a)
and $E/A-\langle NM\mid v_\pi\mid NM\rangle -17\,$MeV (curve b), all obtained
from~\cite{AmTj}.
The two curves have practically the same density dependence verifying that OPEP
contributes little to the saturation mechanism. The subtraction of $17$Mev in
curve (b) makes the scale more compact.

The above results can  be understood qualitatively by examining   the  2nd
order contributions to the $G$-matrix.
Keeping only the positive energy $\Ms$ state contributions in the
intermediate states we have

\ber
\langle \v \ell'\mid \b G && \mid  \v \ell\rangle = \langle \v
\ell'\mid\b V\mid\v \ell\rangle +\sum_{\lambda,i}\int\frac{d^3l''}{(2\pi)^3}
\nonumber\\
&&
\times \langle \v \ell'\mid V\mid\v l''\rangle\frac{\b Q_{Pauli}} {\b
W_0-\v L^{\,'2}/4-\v l''^{\,2}}\langle \v l''\mid\b V \mid\v
\ell\rangle,
\l{BGE3}
\eer
where $\b W_0= W_0/\Ms $ and $\v L = (\v p_1 +\v p_2)/\Ms$.
Now the tensor force contributes only to the second term
of Eq.~(\ref{BGE3}). It has the form
\ber
\ll {\bar V}_{\pi}\rl= -(\frac{g_{\pi NN}}{2})^2 (\frac{\Ms}{M})^2 
[\ubl \gamma_5 (\ell\!\!\!/'-\ell\!\!\!/)\ul]^{(1)}
\nonumber\\
\times \frac{1}{(\v \ell'-\v
\ell)^2+{\bar m}_\pi^2}[\ubml \gamma_5
(\ell\!\!\!/'-\ell\!\!\!/)\uml]^{(2)} ,
\l{pv}
\eer
where $u$'s are the positive energy spinors. Similarly as in the
non-relativistic
case the 2nd order pion contribution is strongly density dependent because of the Pauli
blocking. However as is exhibited in Eq.~(\ref{pv}) in the relativistic case
the effective coupling of the pion to the nucleon is suppressed by a
factor of $\frac{M^*}{M}$.

The $\frac{M^*}{M}$ suppresssion can be corroborated in more detail by
the following calculation.  Let us modify the $S$
from the self-consistent calculation of Ref.~\cite{AmTj} by multiplying it with
the factor $\alpha\leq 1$ thus generating a $M^*=M+\alpha S$.  By  using the
modified scalar self-energy in the nucleon propagators we recalculate first the
$G$ matrices and then $E/A$ and finally $\lNM V_{\pi}\rNM$ using
Eq.~(\ref{Vpi}). Only the $\alpha=1$ analysis is self-consistent, others
are not. But such a calculation is particularly suitable to
exhibit the role of $M^*/M$ on the OPEP contribution.
The  Fig.~\ref{vpisoft} exhibits clearly the damping  due to decreasing $M^*/M$.  We stress that the mechanism of damping is generic to any relativistic treatment and not particular to either Ref.~\cite{AmTj} or the use of Ref.~\cite{BSu}.

\begin{figure}[htb]
\epsfxsize=2.5in
\epsfysize=2.5in
\vspace{0.2in}
\epsffile{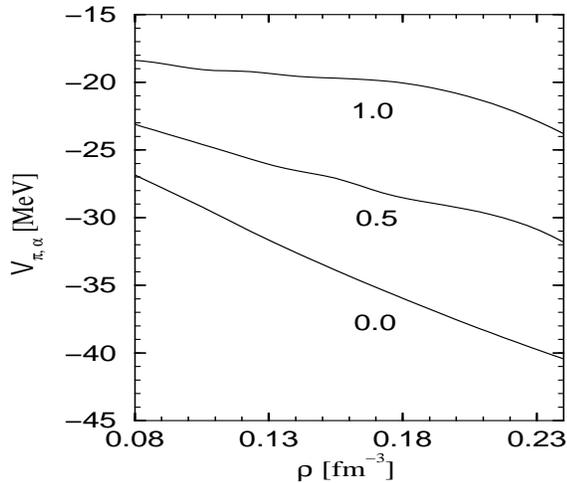}
\vspace{0.2in}
\caption[F4]{Plots of $\langle NM\mid V_\pi\mid NM\rangle$ with the parameters of
Ref.~\protect{\cite{AmTj}}. In the G-matrix calculations $S$ is replaced with
$\alpha S$. The plots are for $\alpha=0.$, $0.5$ and $1.0$. The last one is the
result of  self-consistent calculations of Ref.~\protect{\cite{AmTj}}. The
other two are not  self-consistent.}
\l {vpisoft}
\end{figure}

We want to be careful that the present work not be interpreted as
providing support for MFT. As shown in Fig.~\ref{mfte}, the actual strengths of
the scalar and vector potentials found in the OBE interaction does lead in
a MFT treatment to distinctly different predictions from D-B calculations for
$E/A$. Undoubtedly, if one releases oneself from the constraint
of fitting NN data and  freely chooses the NN interaction one can obtain
proper binding and saturation of nuclear matter with a MFT calculation.

\begin{figure}[htb]
\epsfxsize=2.5in
\epsfysize=2.5in
\vspace{0.2in}
\epsffile{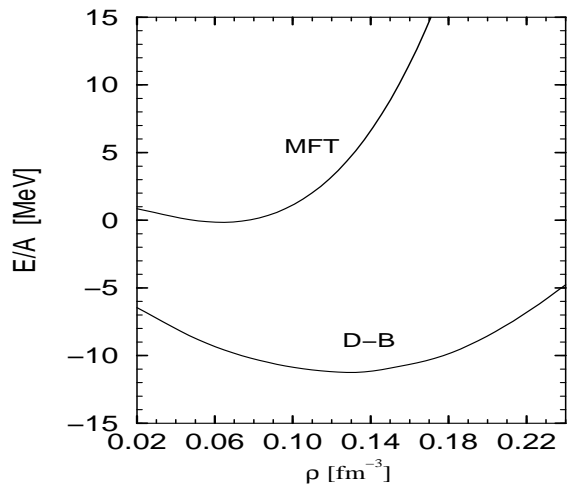}
\vspace{0.2in}
\caption[F4]{Plots of $E/A$ from the D-B calculations of
Ref.~\protect{\cite{AmTj}} and from a MFT calculation with the same
quasipotential. }
\l {mfte}
\end{figure}

In conclusion, we have established that in the relativistic treatment  the  tensor force
contributions are reduced in size in nuclear matter.  Because of this, in
complete contrast to the non-relativistic situation, they cease to play  a role
in the saturation mechanism.
The reduction of the tensor force contributions is principally due to the
relativistic $\Ms/M$ effect, as illustrated with Fig.~\ref{vpisoft}.
But even the reduced role of OPEP is not negligible. As noted,
it contributes $-20$MeV to $E/A$.
This and an enormous body of pre-existing evidence lead us to conclude that there is no viable non-relativistic
theory of nuclear matter based on the use of NN forces
which fit two-body data which is compatible with the relativistic theory.
 The dominant mechanism of
saturation of nuclear matter is basically very different in the two approaches. In the
non-relativistic approach it is the density-dependent reduction due to Pauli blocking of the attraction from tensor force.  while in the relativistic
approach it is the reduction of the rate of growth with increasing $\rho$ of the attraction from the  scalar  field   relative to the  growth of  repulsion from the vector field. 

\vspace{0.5cm}
\noindent
This work was supported in part by DOE Grant DOE-FG02-93ER-40762.
One of us (MKB) thanks the colleagues at the IKP, Forschungszentrum J\"{u}lich, for their hospitality and  A. v. Humboldt
Foundation for an award  for senior U.S. physicist. We thank F. Gruemmer 
for making his nuclear many-body program available to us.


\begin{thebibliography}{100}
\bibitem{FG} This result was generated with a Brueckner theory program of Gruemmer (J\"{u}lich) using the Bonn C potential described in Ref.~\cite{Mach}.
\bibitem{Bethe1}H. A. Bethe, Phys. Rev. {\bf 103}, 1353 (1956);
Ann. Rev. Nucl. Sci. {\bf 21}, 93 (1971).
\bibitem{HumTj} E. Hummel and J.A. Tjon, Phys.
Rev. C {\bf 49}, 21 (1994).
\bibitem{mth} B. ter Haar and R. Malfliet, Phys. Rep. {\bf
149}, 207  (1987).
\bibitem{AmTj} A. Amorim and J. A. Tjon, Phys. Rev. Lett., {\bf 68}, 772
(1992).
\bibitem{Mach}R. Machleidt, Adv. Nucl. Phys. {\bf 19}, 189 (1989).
\bibitem{SW} B. D. Serot and J. D. Walecka, Adv. Nucl. Phys. {\bf 16}, 1
(1986).
\bibitem{BT} M. K. Banerjee and J. A. Tjon, Phys. Rev. {\bf C56}, 497 (1997).
\bibitem{HS} C. Horowitz and B. D. Serot, Nucl. Phys. {\bf A464} (1987) 613.
\bibitem{BSu} R. Blankenbecler and R. Sugar, Phys. Rev. {\bf 142}, 1051 (1966);
A.A. Logunov and A.N. Tavkhelidze, Nuov. Cim. {\bf 29} 380
(1963).
\bibitem{BC} B. C. Clark, R. L. Mercer and P. Schwandt, Phys. Lett. {\bf 122B}
211 (1983).
\bibitem{SJW} J. A. McNeil, J. R. Shepard and S. J. Wallace, Phys. Rev. Lett.,
{\bf 50}, 1439 (1983); {\it ibid} {\bf 50}, 1443 (1983).
\bibitem{SWJT} N. Ottenstein, S. J. Wallace and J. A. Tjon, Phys. Rev. {\bf C
38}, 2272 (1988).
\end{thebibliography}
\end{document}